\documentclass{vgtc}                          

\graphicspath{{figures/}{pictures/}{images/}{./}} 

\usepackage{times}                     

\usepackage{tabu}                      
\usepackage{booktabs}                  
\usepackage{lipsum}                    
\usepackage{mwe}                       
\usepackage{todonotes}

\usepackage{mathptmx}                  

\onlineid{0}

\vgtccategory{Research}

\vgtcinsertpkg

\title{Exploring Human-AI Interaction with Patient-Generated Health Data Sensemaking for Cardiac Risk Reduction}

\author{Pavithren V S Pakianathan\thanks{e-mail: pavithren.pakianathan@lbg.ac.at}\\ %
        \parbox{1.4in}{\scriptsize \centering Ludwig Boltzmann Institute for Digital Health and Prevention\\Salzburg, Austria\\LMU Munich
        \\Munich, Germany}%
\and Rania Islambouli\thanks{e-mail: rania.islambouli@lbg.ac.at}\\ %
     \parbox{1.4in}{\scriptsize \centering Ludwig Boltzmann Institute for Digital Health and Prevention\\Salzburg, Austria}%
\and Hannah McGowan\thanks{e-mail: hannah.mcgowan@lbg.ac.at}\\ %
     \parbox{1.4in}{\scriptsize \centering Ludwig Boltzmann Institute for Digital Health and Prevention\\Salzburg, Austria}
\and Diogo Branco\thanks{e-mail: djbranco@fc.ul.pt}\\ %
     \parbox{1.4in}{\scriptsize \centering LASIGE, Faculdade de Ciências,\\ Universidade de Lisboa\\Lisboa, Portugal}
\and Tiago Guerreiro\thanks{e-mail: tjguerreiro@ciencias.ulisboa.pt}\\ %
     \parbox{1.4in}{\scriptsize \centering LASIGE, Faculdade de Ciências,\\ Universidade de Lisboa\\Lisboa, Portugal}
\and Jan David Smeddinck\thanks{e-mail: jan.smeddinck@lbg.ac.at}\\ %
     \parbox{1.4in}{\scriptsize \centering Ludwig Boltzmann Institute for Digital Health and Prevention\\Salzburg, Austria}
     }

\teaser{
  \centering
  \includegraphics[width=0.7\linewidth]{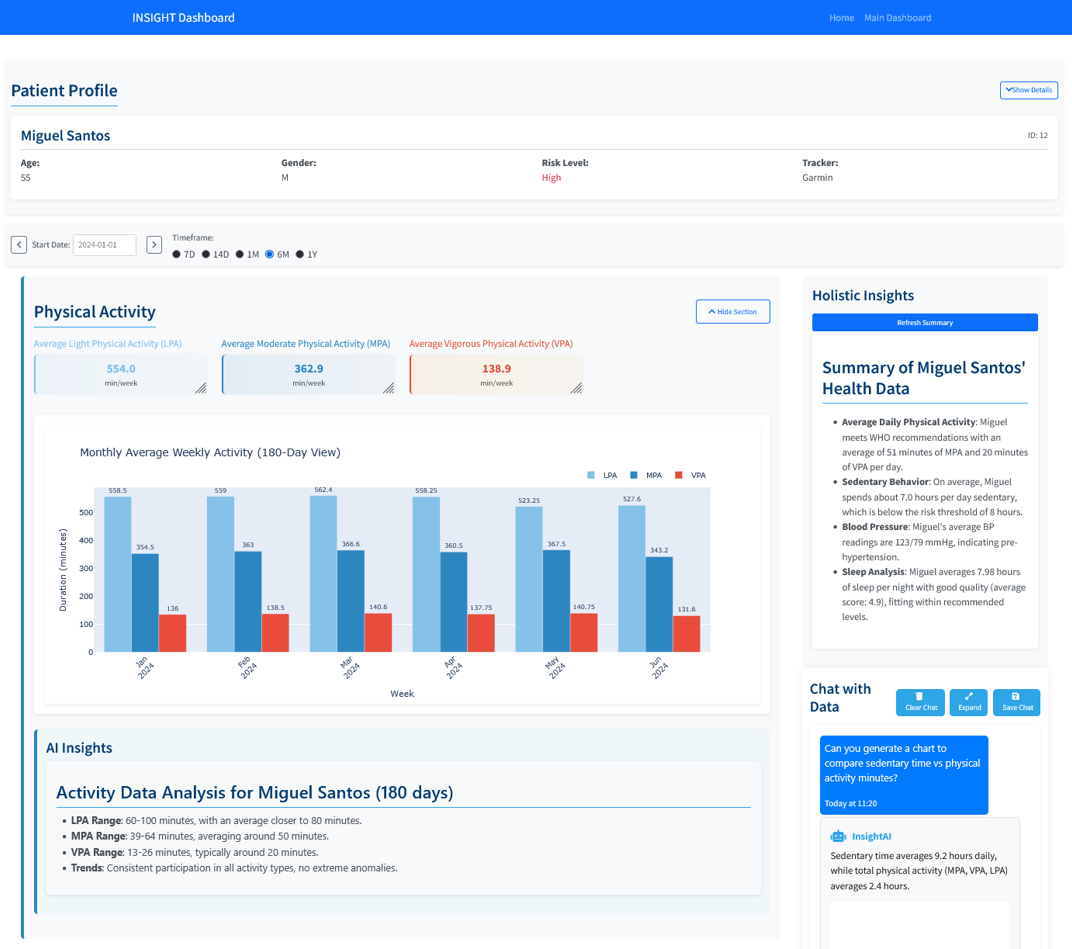}
  \caption{A snapshot of the INSIGHT Dashboard's user interface. 
For a complete view of the dashboard, refer to \textbf{Supplement 1}.
}
  \label{fig:dashboard}
}

\abstract{
Patient-generated health data (PGHD) allows healthcare professionals to have a holistic and objective view of their patients. However, its integration in cardiac risk reduction remains unexplored. Through co-design with experienced healthcare professionals (n=5) in cardiac rehabilitation, we designed a dashboard, INSIGHT (\textbf{IN}vestigating the potential\textbf{S} of Pat\textbf{I}ent \textbf{G}enerated Health data for CVD Prevention and Re\textbf{H}abili\textbf{T}ation), integrating multi-modal PGHD to support healthcare professionals in physical activity planning in cardiac risk reduction. To further augment healthcare professionals’ (HCPs’) data sensemaking and exploration capabilities we integrate large language models (LLMs) for generating summaries and insights and for using natural language interaction to perform personalized data analysis. The aim of this integration is to explore the potential of AI in augmenting HCPs’ data sensemaking and analysis capabilities.
} 

\keywords{cardiac disease, cardiac rehabilitation, patient generated health data, data sensemaking, pervasive health, physical activity planning, multi-modal data.}

\begin{document}
\maketitle


\section{Introduction} 
Multi-modal patient-generated health data from sensors and wearables offer healthcare professionals a holistic and objective view of patients, enabling more personalized care \cite{wendrich_digital_2022}. In the context of cardiac risk reduction, multi-modal PGHD from wearables and sensors such as physical activity, sedentary time, sleep and blood pressure could offer insights into patient vital signs and behaviours and help facilitate physical activity planning \cite{golbus_digital_2023}. 
However integration of such PGHD in clinical workflows to support decision-making has several barriers such as insufficient time, irrelevant data, non-standardization and lack of data science literacy amongst HCPs \cite{westCommonBarriersUse2018m}. In 2023, American Heart Association issued an advisory calling for digital technologies in cardiac rehabilitation to be aligned to clinician's workflows, prioritize automated interpretation of health data using AI to prevent HCPs' data overload and allow for personalised and focussed care for patients \cite{golbus_digital_2023,topol2019deep}. Recent works have investigated how Large-Language Models (LLMs) could augment sense making of fitness, health and medical data \cite{kim_health-llm_2024,li_vital_2025,stromel_narrating_2024,scholich_augmenting_2024}. However such integration in the context of cardiac risk reduction is unexplored. We co-designed a dashboard with HCPs to facilitate data-enabled physical activity planning workflows and aim to explore AI integration to augment data-sensemaking and exploration.

\section{Method}
Through a multi-stage design process with HCPs in cardiovascular rehabilitation (n=5), we co-designed the INSIGHT dashboard, a work-in-progress tool, which integrates PGHD and facilitates HCPs’ decision-making during physical activity planning. 
Firstly, in a situated, 2-week study, six healthy participants self-tracked physical activity and vital signs using a smartwatch, blood pressure monitor and weighing scale and discussed their data with two healthcare professionals in a physical activity planning consultation session. We observed the consultations and interviewed HCP to identify challenges and enablers and to understand how PHGD could be integrated into workflows.

Next, a card-sorting workshop \cite{roy_card-based_2019} was conducted with two cardiac rehabilitation HCPs to validate initial findings and narrow down data and sensemaking needs that align with their workflows. This qualitative method is useful in stimulating idea generation \cite{roy_card-based_2019} and the  use of cards help researchers understand the mental models of participants, informing the design of information systems \cite{serban_i_2023}. 

Using the findings from previous steps, we designed an initial low-fidelity mock-up to support a data driven clinical workflow in physical activity planning. We conducted another round of feedback session with the initial HCP participants and another senior HCP using the mock-ups of the system to refine the feedback. Finally, using all the feedback from the HCPs, the initial version of the ~INSIGHT dashboard was developed.

\section{INSIGHT Dashboard}
\begin{figure}
    \centering
    \includegraphics[width=\linewidth]{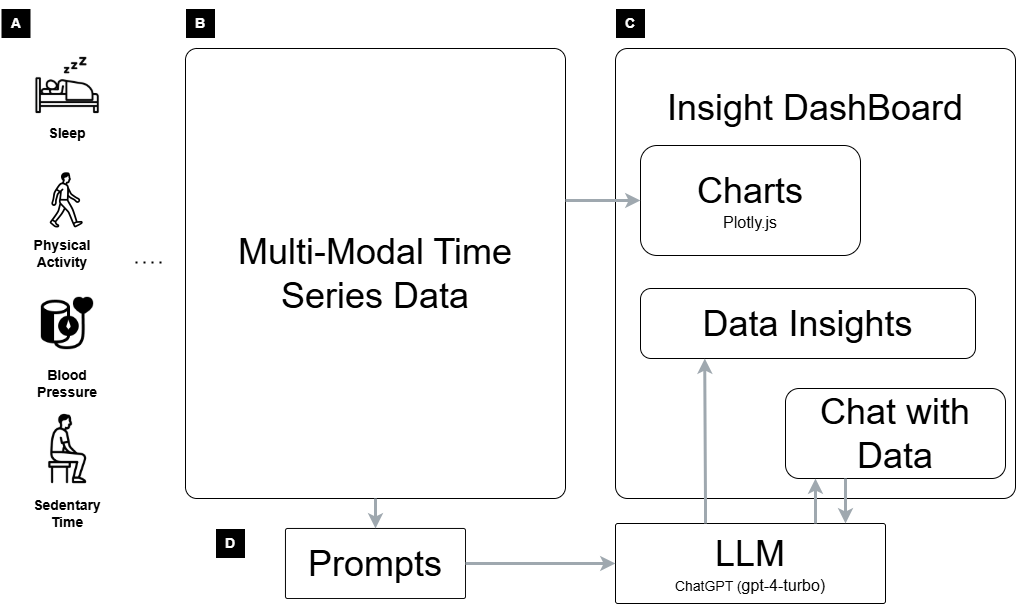}
    \caption{Diagram illustrating the system's architecture}
    \label{fig:system}
\end{figure}

The INSIGHT dashboard (Figure \ref{fig:dashboard}), was developed using the Plotly dash framework (refer to Figure \ref{fig:system} for system architecture) and integrates key information such as patient characteristics and risk factors (comorbidities, medication prescription, demographic information), four PGHD modalities: 1) physical activity minutes (light, moderate and vigorous), 2) sedentary time, 3) blood pressure, 4) sleep duration and quality. Furthermore, HCPs can navigate through the data using temporal scrolling choosing between different time frames such as one week to one year. These information were highlighted by HCPs as key variables for decision-making. 

\subsection{LLM Integration}
\begin{figure}
    \centering
    \includegraphics[width=\linewidth]{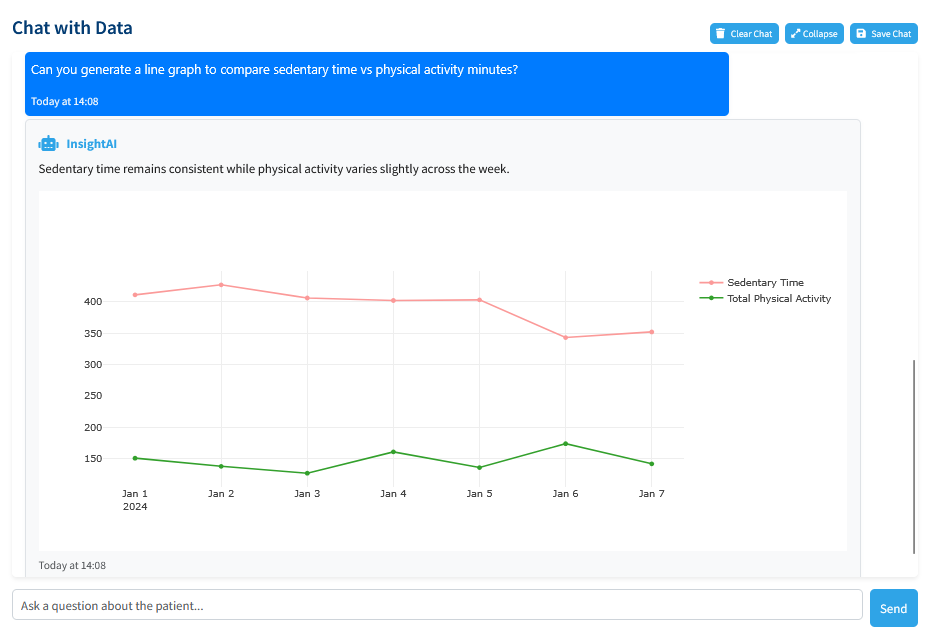}
    \caption{\textit{Chat With Data} Interface showing response to HCP's query asking for a comparison between sedentary time and physical activity minutes}
    \label{fig:chat}
\end{figure}

To  explore how LLMs could augment data sensemaking \cite{vazquez_are_2024} and exploration for HCPs in the context of cardiac risk reduction, we integrated LLM summary and insights capabilities through API integration with ChatGPT. Each chart -  Physical activity, Sedentary time, Blood Pressure and Sleep - has an accompanying natural language summary. Additionally, there is a holistic insight summary capability which provides overall insights from the four data modalities. The summaries are refreshed whenever there is a change in the selected time-range. The prompts were guided by the key information needs previously highlighted by HCPs, including: trends, averages, anomalies, maximum and minimum values, and a comparison with World Health Organization guidelines on physical activity. HCPs can also use natural language to interact with the data, performing personalized analysis and generating interactive charts through an integrated \textit{Chat with Data} feature (Figure \ref{fig:chat}). The aim of this feature is to bridge potential gaps in data science literacy among healthcare professionals and facilitate quick data analysis. All LLM prompts were explicitly crafted to prevent the generation of medical recommendations or exercise prescriptions.

\section{Future work}
Future work will be a empirical study to evaluate the usability of the interface, how it can fit into HCP workflows, whether AI summaries reduces workload and augments data analysis capabilities. Furthermore, we will conduct qualitative investigations with HCPs perceptions and acceptance of AI-integration. We will particularly focus factors that influence trust and reliance, how well AI might fit into their workflows, identify barriers and  enablers, developing risk mitigation strategies for implementation.



\acknowledgments{
The authors wish to thank the healthcare professionals at the Salzburg Landeskrankenhaus (SALK), Salzburg, Austria for their input in designing the system.}

\bibliographystyle{abbrv-doi}

\bibliography{template}
\end{document}